\documentclass[pra,superscriptaddress,twocolumn,showpacs,floatfix]{revtex4}
\usepackage{graphicx}
\usepackage{epstopdf}
\usepackage{amsmath}
\usepackage{bm}

\begin{document}
\title{Observation of St\"{u}ckelberg oscillations in accelerated optical lattices}

\author{A. Zenesini}
\affiliation{Institut f\"ur Experimentalphysik, University of Innsbruck, Technikerstrasse 25, 6020 Innsbruck, Austria}

\author{D. Ciampini}
\affiliation{CNISM UdR Pisa, Dipartimento di Fisica {\it E. Fermi},
Universit\`{a} di Pisa, Largo Pontecorvo 3, I-56127 Pisa, Italy}
\affiliation{INO-CNR, Dipartimento di Fisica {\it E. Fermi},
Universit\`{a} di Pisa,  Largo Pontecorvo 3, I-56127 Pisa, Italy}

\author{O. Morsch}
\affiliation{INO-CNR, Dipartimento di Fisica {\it E. Fermi},
Universit\`{a} di Pisa,  Largo Pontecorvo 3, I-56127 Pisa, Italy}

\author{E. Arimondo}
\affiliation{CNISM UdR Pisa, Dipartimento di Fisica {\it E. Fermi},
Universit\`{a} di Pisa, Largo Pontecorvo 3, I-56127 Pisa, Italy}
\affiliation{INO-CNR, Dipartimento di Fisica {\it E. Fermi},
Universit\`{a} di Pisa,  Largo Pontecorvo 3, I-56127 Pisa, Italy}

\begin{abstract}
We report the experimental observation of St\"{u}ckelberg oscillations of matter waves in optical lattices. Extending previous work on Landau-Zener tunneling of Bose-Einstein condensates in optical lattices, we study the effects of the accumulated phase between two successive crossings of the Brillouin zone edge. Our results agree well with a simple model for multiple Landau-Zener tunneling events taking into account the band structure of the optical lattice.

\end{abstract}

\pacs{03.65.-w, 67.85.Jk, 03.75.Lm, 03.75.Kk}

\maketitle
St\"{u}ckelberg oscillations arise when a quantum system is forced more than once through an avoided crossing of two of its energy levels \cite{Landau,Zener,Stueckelberg}. Each passage through the avoided crossing leads to Landau-Zener tunneling in which the initial wavefunction is coherently split between the two energy levels, and in the next passage the phase accumulated between the energy levels leads to quantum interference. As a result, the populations in the two energy levels after several Landau-Zener passages are very sensitive to the dynamics between the passages. While
St\"{u}ckelberg oscillations were first predicted more than eighty years ago for atomic collisions, they have received renewed interest in recent years as a tool for the spectroscopic investigation \cite{shevchenko_10} of, e.g., superconducting artificial atoms \cite{Oliver_05,Sillanpaa,Oliver_09}, molecule formation in ultra-cold gases \cite{mark_07}, and Rydberg excitations \cite{ditzhuijzen_09}. Here we report results on the realization of St\"{u}ckelberg oscillations using Bose-Einstein condensates in accelerated optical lattices. From the measurements of the frequency and damping time of the oscillations we are able to deduce the distance between the energy bands of the optical lattice and the momentum spread of the condensate.

\begin{figure}[h]
\centering
\includegraphics[width=0.3 \textwidth]{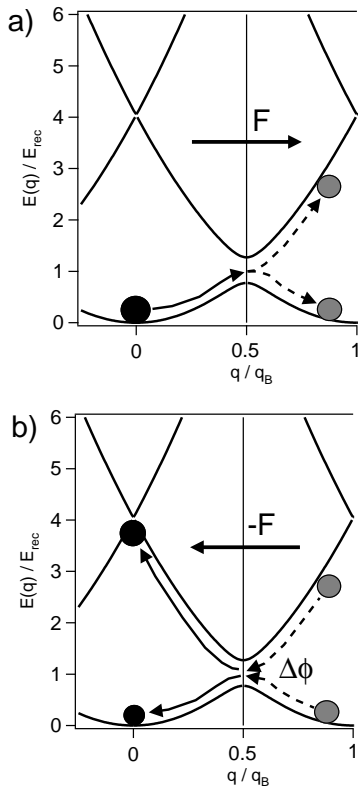}
\caption[1]{Experimental protocol for observing St\"{u}ckelberg oscillations in accelerated optical lattices.}
\label{stueckel_scheme}
\end{figure}

In our experiments we investigate the time evolution of Bose condensates loaded into a spatially periodic potential subjected to an additional (inertial) static force, as described in detail in~\cite{zenesini_09}. The dynamics of ultracold atoms in a tilted optical lattice can be described by the well-known Wannier-Stark Hamiltonian~\cite{tayebirad_10} determining the atomic evolution in an optical lattice of depth $V_0$ and spatial period $d_{\rm L}=\lambda_{\rm L}/2$ determined by the wavelength $\lambda_{\rm L}$  of the laser creating the periodic potential. The characteristic energy scale of the system is the recoil energy $E_{\rm rec}=\pi^{2}\hbar^{2}/2 M d_{\rm L}^{2}$, where $M$ is the atomic mass.  The band-structure representation of the energy $E_n(q)$ of the $n$-th band versus the quasimomentum $q$ is represented in Fig. \ref{stueckel_scheme}. The observation of St\"{u}ckelberg oscillations is based on a sweep of the quasimomentum $q$ given by $q(t)=Ft$ induced by a force $F$ (measured here in units of $E_\mathrm{rec}/d_L$) acting on the atoms in the rest frame of the lattice.

Our experiments can be modeled (neglecting  the finite duration of the Landau-Zener tunneling event \cite{zenesini_09,vitanov_99}) using a two-state ansatz for the two lowest energy bands $E_{0,1}(q)$ in the optical lattice, described by the state vectors $\left(\begin{array}{c}1\\0\end{array}\right)$ and $\left(\begin{array}{c}0\\1\end{array}\right)$. The different steps of the experimental protocol are then represented by matrices $\textsf{M}_i$ acting on the general state vector $\left(\begin{array}{c}a\\b\end{array}\right)$ \cite{asshab_07}:
\begin{itemize}
\item LZ tunneling event:

\[
\textsf{M}_\mathrm{LZ}(P_\mathrm{LZ}) =
\left( {\begin{array}{cc}
 \sqrt{1-P_\mathrm{LZ}}e^{-i(\phi_{\rm St}-\pi/2)} &  -\sqrt{P_\mathrm{LZ}}   \\
  \sqrt{P_\mathrm{LZ}}  &  \sqrt{1-P_\mathrm{LZ}}e^{i(\phi_{\rm St}-\pi/2)}   \\
 \end{array} } \right),
\]

where $P_\mathrm{LZ}=\exp(-\pi/\gamma)$ is the Landau-Zener (LZ) tunneling probability with $\gamma=32Fd_{\rm L}E_{\rm rec}/(\pi V_0^2)$, and $\phi_{\rm St}=\pi/4-\left[1+ln(2\gamma)\right](2\gamma)^{-1}+arg\Gamma[1-(2\gamma)^{-1}i]$ the Stokes phase~\cite{vitanov_99,shevchenko_10}.

\item acceleration from initial quasimomentum $q_\mathrm{ini}$ to final quasimomentum $q_\mathrm{fin}$:

\[
\textsf{M}_\mathrm{acc}(q_\mathrm{ini},q_\mathrm{fin}) =
\left( {\begin{array}{cc}
 e^{-i \Phi_\mathrm{0}(q_\mathrm{ini},q_\mathrm{fin})} &  0   \\
  0  &   e^{-i \Phi_\mathrm{1}(q_\mathrm{ini},q_\mathrm{fin})}   \\
 \end{array} } \right),
\]

where,  using the linear dependence of $q$ on time, $ \Phi_\mathrm{n}(q_\mathrm{ini},q_\mathrm{fin}) = \frac{1}{\hbar F} \int^{q_\mathrm{ini}}_{q_\mathrm{fin}} E_{n}(q) dq$.

\item waiting for time $t_\mathrm{wait}$ at quasimomentum $q$:

\[
\textsf{M}_\mathrm{wait}(q,t_\mathrm{wait}) =
\left( {\begin{array}{cc}
 e^{-i E_{0} (q) t_\mathrm{wait}/\hbar} &  0   \\
  0  &   e^{-i E_{1} (q)t_\mathrm{wait}/\hbar}    \\
 \end{array} } \right).
\]
\end{itemize}

A general experiment of crossing the zone edge at $q=0.5\,q_B$, accelerating to $q_\mathrm{fin}$, waiting for time $t_\mathrm{wait}$, accelerating back to $q=0.5\,q_B$ and crossing the zone edge a second time (see Fig. 1) can then be described by the total transfer matrix
\begin{eqnarray}
\textsf{M}_\mathrm{tot}= \textsf{M}^{\dagger}_\mathrm{LZ}\textsf{M}^{\dagger}_\mathrm{acc}\textsf{M}_\mathrm{wait} \textsf{M}_\mathrm{acc}\textsf{M}_\mathrm{LZ},
\end{eqnarray}
where the arguments of the matrices have been omitted for brevity. If the system is initially in the lowest energy band of the lattice, the probability $P_0$ of finding it in that state after the above protocol is
\begin{equation}
P_0=4P_\mathrm{LZ}(P_\mathrm{LZ}-1)\sin^2\left(\frac{\Delta\phi}{2}+\phi_{\rm St}\right)+1,
\end{equation}
where $\Delta\phi=[(\Phi_0-\Phi_1)+(E_0-E_1)t_\mathrm{wait}/\hbar]$ denotes the total relative phase between the two bands accumulated during the acceleration and waiting parts of the protocol.

Our experimental setup is described in detail in \cite{zenesini_09}. Briefly, we create Rb condensates containing up to $10^5$ atoms in a crossed dipole trap and subsequently load them into a one-dimensional optical lattice formed by two linearly polarized counter-propagating gaussian beams far-detuned from the Rb atomic resonance. The dipole trap beam perpendicular to the lattice direction is then switched off and the lattice is accelerated by applying appropriate frequency chirps on the radio-frequency generators powering the two acousto-optic modulators used to control the power in the lattice beams. Finally, the dipole trap beam and the lattice beams are switched off and an absorption image is taken after $23\,\mathrm{ms}$ of time-of-flight, which allows us to deduce the survival probability in the ground state $P_0=N_0/N_\mathrm{tot}$ from the relative number of atoms in the $p=0$ momentum class.

\begin{figure}[ht]
\centering
\includegraphics[width=0.5 \textwidth]{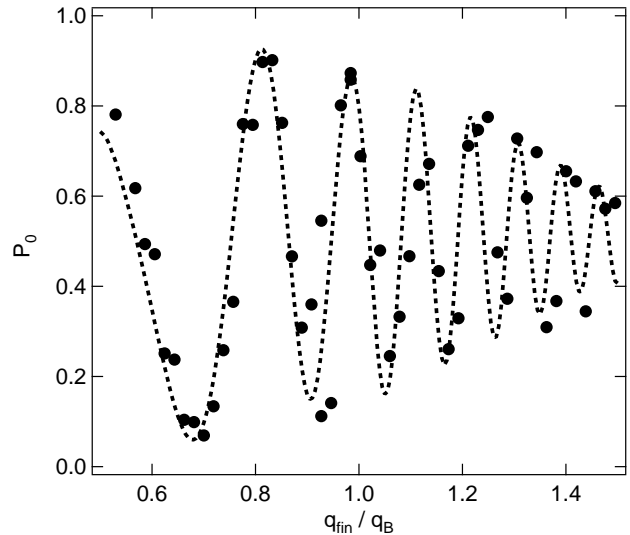}
\caption[1]{St\"{u}ckelberg oscillations after two crossings of the zone edge in an optical lattice. Shown here is the survival probability in the lowest energy band $P_0$ for $V_0=1.4\,E_\mathrm{rec}$ and $F=1.197\,E_\mathrm{rec}/d_L$ as a function of $q_\mathrm{fin}$; for these experimental parameters the Stokes phase is $\phi_{\rm St}\approx -\pi/3$. The dashed line is a numerical simulation assuming a momentum spread of the condensate of $\Delta q=0.03q_B$. The data points are the results of single experiments and hence no error bar is reported; shot-to-shot fluctuations for constant experimental parameters are around $10\%$.}
\end{figure}

St\"{u}ckelberg oscillations are then observed by first loading the condensate adiabatically into the lowest energy band of a lattice ($V_0=1.4\,E_\mathrm{rec}$) with zero quasimomentum (stationary lattice), accelerating the lattice across the edge of the Brillouin zone at $q=0.5\,q_{\rm B}$ (where $q_{\rm B}=2\pi\hbar/d_{\rm L}$ is the Bloch momentum), reversing the sign of the acceleration, crossing the edge of the Brillouin zone a second time and finally stopping at zero quasimomentum. The results of such an experiment with $t_\mathrm{wait}=0$ (i.e. $\textsf{M}_\mathrm{wait}=\textsf{I}$) and varying $q_\mathrm{fin}$ are shown in Fig. 2. Since the energy separation between the bands varies (approximately) linearly with $q_\mathrm{fin}$, the relative phase $\Delta\phi$ is a quadratic function of $q_\mathrm{fin}$ and hence the frequency of the oscillations varies, leading to `chirped' St\"{u}ckelberg oscillations. Beyond quasimomentum $q=q_B$ in Fig. 1 our simple two-level model is, strictly speaking, no longer valid as avoided crossings with higher bands will lead to a loss of population from the excited band. Since the tunneling probability to those higher bands (and back) is close to unity, however, the dominant correction to the two-level approximation will be in the phase factor.

In spite of the simplifying assumptions, the agreement with our simple model (using Mathieu-function solutions for the band structure of the lattice) is still very good. From the total acceleration time $t_\mathrm{acc}=0.4\,\mathrm{ms}$ between successive crossings of the zone edge for the maximum $q_\mathrm{fin}$, the damping time of the St\"{u}ckelberg oscillations of Fig. 2 is found to be around $\tau=0.35\,\mathrm{ms}$. We can account reasonably well for this damping by assuming a momentum spread of the condensate of $\Delta q=0.03q_B$ in our simulations, which in practice is partly due to the intrinsic momentum width of the condensate and partly to dephasing mechanisms such as dynamical instabilities that occur during the acceleration phase \cite{cristiani_04}. The momentum spread leads to an accumulation of relative phase of different parts of the condensate at a different rate depending on their position in the Brillouin zone. We can, therefore, connect the momentum spread $\Delta q$ to the spread in the accumulated relative phase $\Delta \phi$ in the following way: Assuming that the two lowest energy bands for a shallow lattice correspond roughly to the free particle dispersion relations $E_n(q)=(q-nq_B)^2/(2M)$ (except around $q=0.5\,q_B$), we find $d\Delta\phi/dq=8E_\mathrm{rec}T_\mathrm{tot}/\hbar$ with $T_\mathrm{tot}=2t_\mathrm{acc}+t_\mathrm{wait}$. For a given momentum spread $\Delta q$, the damping time $\tau$ of the St\"{u}ckelberg oscillations is then $\tau=\hbar/(8E_\mathrm{rec}(\Delta q/q_B))$, giving $\Delta q/q_B=0.02$ for $\tau=0.35\,\mathrm{ms}$, in good agreement with the simulations.

\begin{figure}[ht]
\centering
\includegraphics[width=0.5 \textwidth]{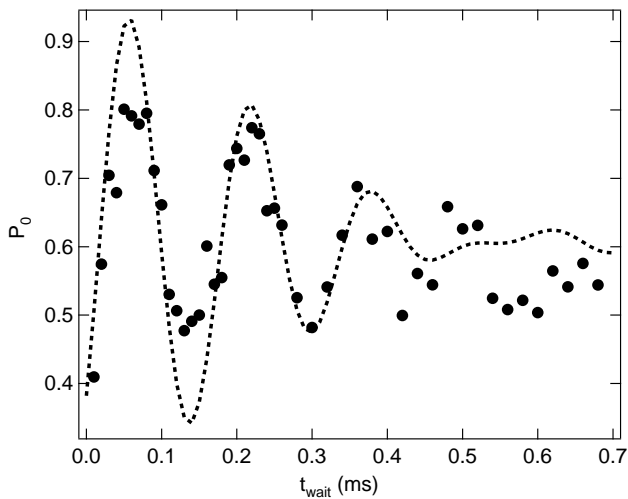}
\caption[2]{Survival probability in the lowest energy band $P_0$ for $V_0=1.1\,E_\mathrm{rec}$ and $F=1.197\,E_\mathrm{rec}/d_L$ as a function of the waiting time at $q_\mathrm{fin}=0.719\,q_B$.}
\end{figure}

We now fix $q_\mathrm{fin}=0.719\,q_B$ and modify the above protocol by inserting a variable waiting time between the two acceleration stages. As can be seen in Fig. 3, this gives rise to St\"{u}ckelberg oscillations with constant frequency since $\Delta\phi$ now varies linearly with $t_\mathrm{wait}$. The measured frequency of the oscillations of around $6.7\,\mathrm{kHz}$ is in good agreement with the expected band separation of $7.07\,\mathrm{kHz}$ at $q=0.719\,q_B$. As in Fig. 2, the oscillations are damped on a timescale of $\approx 0.4\,\mathrm{ms}$.

\begin{figure}[ht]
\centering
\includegraphics[width=0.5 \textwidth]{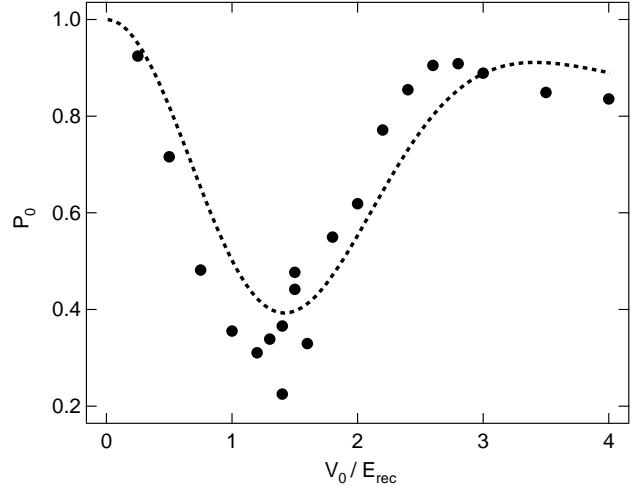}
\caption[3]{Survival probability in the lowest energy band $P_0$ as a function of $V_0$ for $F=1.197\,E_\mathrm{rec}/d_L$, a fixed final quasimomentum $q_\mathrm{fin}=0.757\,q_B$ and $t_\mathrm{wait}=0$. The Stokes phase varies between $-\pi/2$ for $V_0\approx 0.5\,E_\mathrm{rec}$ and $\pi/9$ for $V_0\approx 4\,E_\mathrm{rec}$.}
\end{figure}

Finally, we perform an experiment in which the lattice depth $V_0$ is varied keeping $q_\mathrm{fin}=0.757\,q_B$ and $t_\mathrm{wait}=0$ constant. In this case, both $\textsf{M}_\mathrm{LZ}$ (through $P_\mathrm{LZ}$) and $\textsf{M}_\mathrm{acc}$ (through the band structure and hence $E_{0,1}(q)$) are varied and the resulting variation in $P_0$ depends sensitively on the exact parameters of the system. In the two extreme cases of very small and very large lattice depth $V_0$, however, the experiment has an intuitively predictable outcome (which can also be read off immediately from Eq. 2). For small $V_0$, the Landau-Zener tunneling probability $P_{LZ}\approx 1$ and hence complete tunneling will occur in both Landau-Zener passages, so the system ends up in its initial state after the above protocol. For large $V_0$, we also expect the system be in the ground state energy band after two Landau-Zener passages, but now this is because $P_{LZ}\approx 0$ and hence tunneling into the other band is strongly suppressed, so the system stays adiabatically in the ground state at all times.

Fig. 4 shows the results of the above experiment. As expected, for very large and very small values of $V_0$, $P_0$ is close to $1$, whereas in between $P_0$ reaches a minimum of $\approx 0.3$ for $V_0\approx 1.4\,E_\mathrm{rec}$. We note here that, as can be seen from Eq. 2, the exact behaviour of $P_0$ depends crucially on how $\Delta \phi$ (which includes the Stokes phase) depends on $V_0$. Indeed, if no relative phase were accumulated between the two crossings of the zone edge, we would expect $P_0=1$ independently of the value of $P_{LZ}$.

In summary, we have observed St\"{u}ckelberg oscillations of Bose-Einstein condensates in optical lattices using different experimental protocols. The sensitivity of those oscillations to accumulated phases in the energy bands suggests the use of St\"{u}ckelberg protocols for studying dephasing and related phenomena in optical lattices, for instance in a superlattice~\cite{kling_10}. Our methods for the determination of dephasing and energy splitting may be also be applied to other systems that can be described by a two-state model.

Financial support by the E.U.-STREP "NAMEQUAM" and by a CNISM "Progetto Innesco 2007" is gratefully acknowledged. We thank J. Radogostowicz for assistance.

\bibliographystyle{apsrmp}

\end{document}